\documentclass[twocolumn,showpacs,superscriptaddress,nofootinbib,prl]{revtex4}
\usepackage{latexsym}
\usepackage{bm}
\usepackage{amssymb}
\usepackage{graphicx}

\newcommand{\Dslash}{\ensuremath{{D\kern -0.65em /}}}
\newcommand{\half}{\ensuremath{{\textstyle\frac{1}{2}}}}

\begin{document}

\preprint{\parbox[t]{15em}{\raggedleft
FERMILAB-PUB-04/349-T \\
hep-lat/0411027}}

\title{\boldmath 
Mass of the $B_c$ Meson in Three-Flavor Lattice QCD}

\author{Ian F. Allison}
\author{Christine T.H. Davies}
\affiliation{Department of Physics and Astronomy, Glasgow University,  
	Glasgow, Scotland, United Kingdom}
\author{Alan Gray}
\affiliation{Department of Physics, The Ohio State University,  
	Columbus, Ohio, USA}
\author{Andreas S. Kronfeld}
\author{Paul B. Mackenzie}
\author{James N. Simone}
\affiliation{Theoretical Physics Department, Fermi National
	Accelerator Laboratory,  Batavia, Illinois, USA}
\collaboration{HPQCD, Fermilab Lattice, and UKQCD Collaborations}

\date{November 19, 2004}

\begin{abstract}
We use lattice QCD to predict the mass of the $B_c$ meson.
We use the MILC Collaboration's 
ensembles of lattice
gauge fields, which have a quark sea with two flavors 
much lighter than a third. 
Our final result is $m_{B_c}=6304\pm12^{+18}_{-\;\,0}~\text{MeV}$.
The first error bar is a sum in 
quadrature of statistical and 
systematic uncertainties, and the second is an estimate of heavy-quark
discretization effects.
\end{abstract} 


\maketitle


Recently there has been a significant breakthrough in numerical lattice
calculations of QCD~\cite{Davies:2003ik}.
With new, improved techniques for incorporating light sea quarks,
lattice QCD agrees with experiment at the few percent level for a wide
variety of quantities.
This progress suggests that lattice QCD could play a big role in
particle physics, especially as an aid to understanding the flavor
sector of the Standard Model~\cite{Bernard:2002bk}.

In flavor physics, the central aim is to search for evidence of new
phenomena.
Before applying results from numerical lattice QCD for such purposes,
it is helpful to have as many tests as possible.
Although lattice gauge theory has a solid mathematical foundation,
numerical simulations are not simple.
The impressive results of Ref.~\cite{Davies:2003ik} have been achieved 
only with the fastest method for simulating light quarks.
The price for speed is an unproven assumption (discussed below), which
clearly warrants further scrutiny.
In addition, the cutoff effects of heavy quarks are controlled using
effective field theories.
Although most heavy-quark phenomenology relies on this framework, it is
important to find out how well it describes discretization errors in
lattice calculations.

The ideal way to test a theoretical technique is to predict a
mass or decay rate that is not well-measured experimentally, but will
be measured precisely soon.
Some examples are in leptonic and semileptonic decays of
charmed mesons, which are being measured in the CLEO-$c$ experiment.
They are sensitive to both the light-quark \emph{and} heavy-quark
methods, and are under investigation~\cite{Wingate:2003gm,Aubin:2004ej}.

Another example, pursued here, is the mass of the pseudo\-scalar $B_c$
meson, the lowest-lying bound state of a bottom anti-quark ($\bar{b}$)
and a charmed quark~($c$).
The $B_c$ mass principally tests the heavy-quark methods of lattice QCD.
Based on experience with $\bar{b}b$~\cite{Gray:2002vk} and
$\bar{c}c$~\cite{diPierro:2003bu} mass splittings, we expect only mild
sensitivity to the light quark mass (of the sea quarks) once the mass is
small enough to allow uninhibited creation and annihilation of virtual
light quark pairs.
Preliminary versions of this work have been given at
conferences~\cite{Allison:2004hy}.

Until now, $B_c$ has been observed only in the semileptonic decay
$B_c^+\to J/\psi\,l^+\nu_l$, with a mass
resolution of around $400~\text{MeV}$ \cite{Abe:1998wi,D0:2004pl}.
During Run~2 of the Fermilab Tevatron, $B_c$ is expected be observed in
non-leptonic decays, with a mass resolution estimated to be
$20$--$50~\text{MeV}$ \cite{Anikeev:2001rk}.
Our total uncertainty is much smaller than the current experimental
accuracy, and comparable to the projections, so we may claim to be
predicting the mass of the $B_c$~meson.

Heavy-quark discretization effects are a challenge, because feasible
lattice spacings~$a$ are about the same as the Compton wavelength of the
bottom and charmed quarks.
The distances are both shorter than the typical distance of QCD,
which is about $1~\text{fm}$.
The obvious strategy is to use effective field theories to separate
long- and short-distance scales.
This reasoning has led to the development of non-relativistic QCD
(NRQCD) for quarkonium~\cite{Lepage:1987gg} and heavy-quark effective
theory (HQET) for heavy-light mesons~\cite{Eichten:1987xu}.
In lattice gauge theory, this reasoning has led to two systematic
methods for discretizing the heavy-quark Lagrangian:
lattice NRQCD~\cite{Lepage:1987gg,Lepage:1992tx} and the Fermilab
heavy-quark method~\cite{El-Khadra:1996mp,Kronfeld:2000ck}.
A~strength of both is that the free parameters of the lattice
Lagrangian can be fixed with quarkonium.
Then, with no free parameters, one obtains results for heavy-light
systems (such as $D$ and $B$ mesons).
The same procedure applies here: we obtain $m_{B_c}$ with the same bare
quark masses that reproduce the bottomonium~\cite{Gray:2002vk} and
charmonium~\cite{diPierro:2003bu} spectra.

It is beyond the scope of this Letter to review the \nolinebreak
details of
heavy quarks in lattice gauge theory~\cite{Kronfeld:2003sd}.
The couplings of the Lagrangian are adjusted
so that~\cite{Kronfeld:2000ck}
\begin{eqnarray}
	\mathcal{L}_{\text{lat}} &\doteq & \mathcal{L}_{\text{QCD}} +
		\delta m(\bar{h}^+h^+ + \bar{h}^-h^-) + \nonumber \\ & & 
		\sum_n a^{s_n} f_n(m_Qa) \mathcal{O}_n
	\label{eq:EFT}
\end{eqnarray}
where $\doteq$ can be read ``has the same mass spectrum as.''
The $\delta m$ term is an unimportant overall shift in the mass spectrum;
$h^+$ ($h^-$) is a effective field for quarks (anti-quarks);
the $\mathcal{O}_n$ are the effective operators of the heavy-quark
expansion, of dimension $\dim\mathcal{O}_n=4+s_n$, $s_n\geq1$;
and $a$~is the lattice spacing.
The coefficients $f_n$ arise from the short-distance mismatch between
lattice gauge theory and continuum QCD.
By choosing an \emph{improved} lattice
Lagrangian~$\mathcal{L}_{\text{lat}}$, the $f_n$ can be reduced.
In practice, however, one must vary $a$ and also estimate the effects of
the leading $\mathcal{O}_n$ on the mass spectrum.

Our calculation employs an idea from a quenched
calculation~\cite{Shanahan:1999mv} (omitting sea quarks), namely to use
lattice NRQCD for the $b$~quark and the Fermilab method for the
$c$~quark.
The lattice NRQCD Lagrangian~\cite{Lepage:1992tx} has a better treatment
of interactions of order $v^4$, where $v$ is the heavy-quark velocity.
The Fermilab Lagrangian~\cite{El-Khadra:1996mp} has a better treatment
of higher relativistic corrections, which is helpful since the velocity
of the $c$ quark in $B_c$ is not especially small, $v_c^2\approx0.5$.
Thus, we expect this combination to control discretization effects well.
This choice also means that our calculation directly tests the
heavy-quark Lagrangians used in Ref.~\cite{Davies:2003ik}.

We work with ensembles of lattice gauge fields from the MILC
Collaboration~\cite{Bernard:2001av}.
Each ensemble contains several hundred lattice gauge fields, so
statistical errors are a few per cent.
The gluon fields interact with a sea of ``$2+1$'' quarks: one with
mass~$m_s$ tuned close to that of the strange quark, and the other two
as light as possible.
In this work we use ensembles with light mass $m_l=0.1m_s$,
$m_l=0.2m_s$, and $m_l=0.4m_s$.
The gluon and sea-quark Lagrangians are improved to reduce discretization
effects.
We use three lattice spacings, $a\sim\frac{1}{11}$, $\frac{1}{8}$,
$\frac{2}{11}~\text{fm}$.
Further details are in the MILC Collaboration's 
papers~\cite{Bernard:2001av}.

A drawback of the MILC ensembles is that the sea quarks are incorporated
with ``staggered'' quarks.
A single staggered quark field leads to four species, or ``tastes,'' in
the continuum limit.
Sea quarks are represented (as usual) by the determinant of the
staggered discretization of the Dirac operator.
To simulate 2~tastes (1~taste), the square root (fourth root) of the
4-taste determinant is taken.
The validity of this procedure is not yet proven for lattice QCD,
although a proof does go through in at least one (non-trivial)
context~\cite{Adams:2003rm}.
Moreover, one finds that interacting improved staggered fields split
into quartets~\cite{Follana:2004sz}, as is necessary.
Since our prediction of the $B_c$ mass tests this ingredient of the
calculation (albeit indirectly), we do not assign a numerical error bar
to this issue.

As in Ref.~\cite{Shanahan:1999mv}, we calculate mass splittings,
namely
\begin{eqnarray}
	\Delta_{\psi\Upsilon} & = & 
		m_{B_c} - (\bar{m}_\psi + m_\Upsilon)/2,
	\label{eq:QQ-baseline} \\
	\Delta_{D_sB_s} & = & m_{B_c} - (\bar{m}_{D_s} + \bar{m}_{B_s}),
	\label{eq:sQ-baseline}
\end{eqnarray}
where
$\bar{m}_\psi  = (m_{\eta_c} + 3 m_{J/\psi})/4$,
$\bar{m}_{D_s} = (m_{D_s} + 3 m_{D_s^*})/4$, and
$\bar{m}_{B_s} = (m_{B_s} + 3 m_{B_s^*})/4$
are spin-averaged masses.
We refer to $(\bar{m}_\psi+m_\Upsilon)/2$ and
$(\bar{m}_{D_s}+\bar{m}_{B_s})$ as the ``quarkonium'' and
``heavy-light'' baselines, respectively.
Our result for $m_{B_c}$ comes from our calculated
$a\Delta_{\psi\Upsilon}$ and $a\Delta_{D_sB_s}$ (in lattice units),
combined with the lattice spacing~$a$ and the experimental measurements
of the baselines.
We use the 2S--1S splitting of bottomonium to define $a$, but on the
MILC ensembles several other observables would serve equally
well~\cite{Davies:2003ik}.


Many uncertainties cancel in mass splittings.
Lattice calculations integrate the QCD functional integral with a
Monte Carlo method, and the ensuing statistical error largely cancels
when forming a difference.
The mass shifts $\delta m$ in Eq.~(\ref{eq:EFT}) drop out.
The spin-averaging cancels the contribution of the hyperfine operator
$\bar{h}^{\pm}i\bm{\Sigma}\cdot\bm{B}h^{\pm}$.
(We do not spin-average $\Upsilon$ with $\eta_b$, because the latter
remains unobserved.)
The discretization errors from further terms in Eq.~(\ref{eq:EFT})
cancel to some extent, especially with the quarkonium baseline.
Most crucially, all masses in Eqs.~(\ref{eq:QQ-baseline})
and~(\ref{eq:sQ-baseline}) are ``gold-plated''~\cite{Davies:2003ik}, in
the sense that the hadrons are stable and not especially sensitive to
light quarks.
(Hence we use $D_s$ and $B_s$, not $D$ and~$B$.)

We turn now to a discussion of our numerical work.
First we discuss briefly how to compute the meson masses.
Then we consider systematic effects that can be addressed directly by
varying the bare quark masses (light and heavy).
Finally, we consider the remaining discretization effects, by changing
the lattice spacing and by studying the corrections in
Eq.~(\ref{eq:EFT}).

In lattice QCD, each meson mass is extracted from a two-point
correlation function, which contains contributions from the desired
state and its radial excitations.
We use constrained curve fitting~\cite{Lepage:2001ym}, usually including
5 states, but checking the results with 2--8 states in the fit.
We find that the extraction of the raw masses is straightforward on
every ensemble.

Statistical errors are obtained with the bootstrap method.
The statistical precision on $\Delta_{\psi\Upsilon}$ is about 4\% and on
$\Delta_{D_sB_s}$ about~1.5\%.
But since $\Delta_{\psi\Upsilon}\approx 40~\text{MeV}$ and
$\Delta_{D_sB_s}\approx-1200~\text{MeV}$, the statistical error on
$m_{B_c}$ ends up being much larger with the heavy-light baseline.

\begin{figure}
	\centering
	\includegraphics[width=0.47\textwidth]{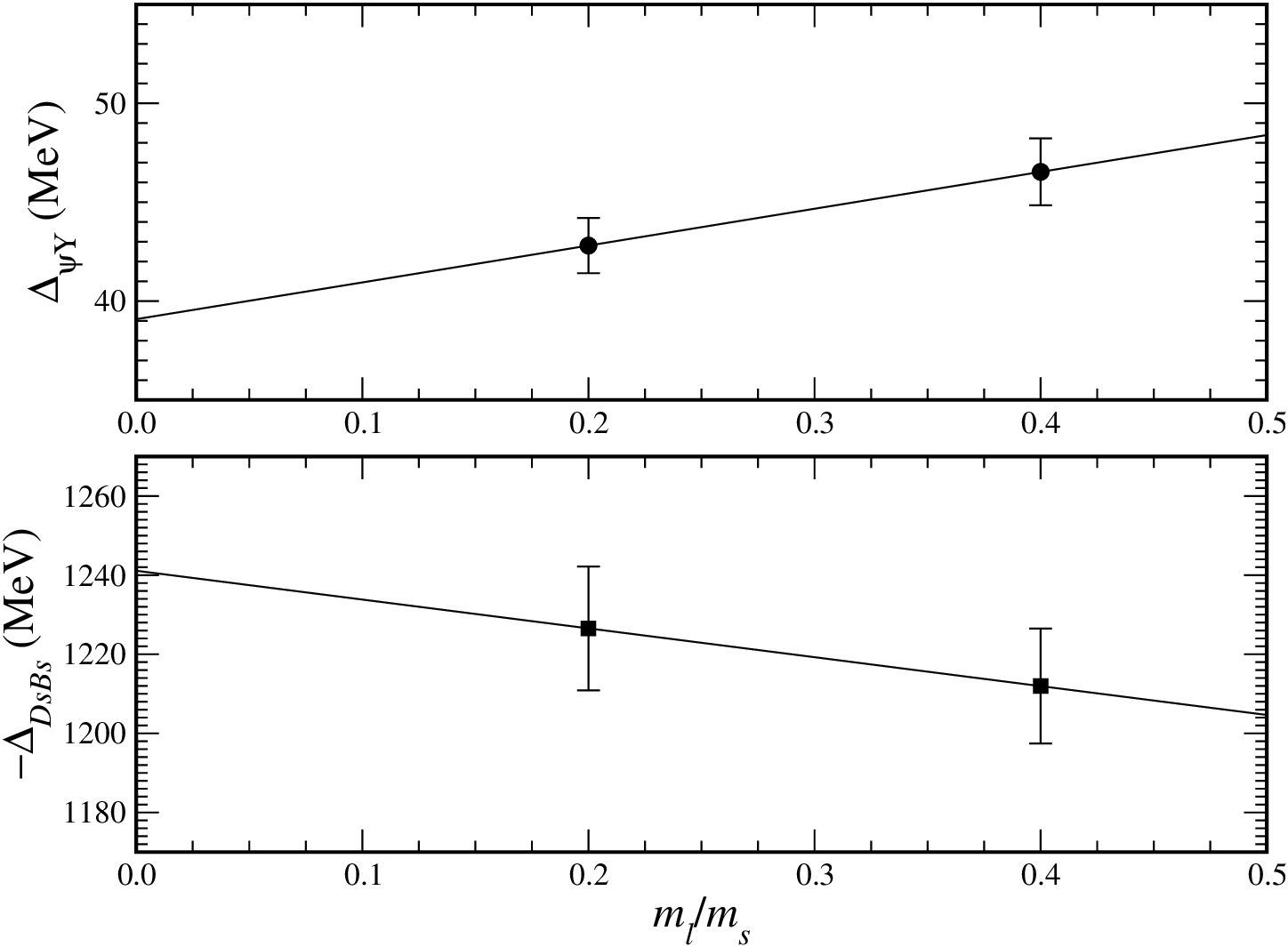}
	\caption{Sea-quark mass dependence of $\Delta_{\psi\Upsilon}$ and
	$\Delta_{D_sB_s}$.}
	\label{fig:chiral}
\end{figure}

Figure~\ref{fig:chiral} shows how the splittings depend on the light
quark mass $m_l$, for the ensembles with
$a\approx\frac{1}{8}~\text{fm}$.
The dependence on $m_l$ is hardly significant.
We extrapolate linearly in $m_l/m_s$, down to the value that reproduces
the pion mass~\cite{Bernard:2002bk}.
The mild dependence on $m_l$ also suggests that the uncertainty from the
known (but small) mistuning of the strange quark sea is completely
negligible.

The bare masses of the heavy quarks are chosen as follows.
Since the overall mass is shifted [by $\delta m$ in Eq.~(\ref{eq:EFT})],
we compute the kinetic energy of $\bar{b}b$ and $\bar{c}c$ mesons of
(small) momentum $\bm{p}$, and choose the bare $b$ and $c$ quark masses
so that it is $\bm{p}^2/2m$, where $m$ is the physical $\bar{Q}Q$ mass.
The statistical and systematic uncertainties of the kinetic energy imply
a range of bare quark masses.
We compute the effect on $B_c$ for different bare $b$ and $c$ masses and
derive an error of $10~\text{MeV}$ ($5~\text{MeV}$) in
$\Delta_{\psi\Upsilon}$ and $\Delta_{D_sB_s}$ from this source.

\begin{figure}[b]
	\centering
	\includegraphics[width=0.47\textwidth]{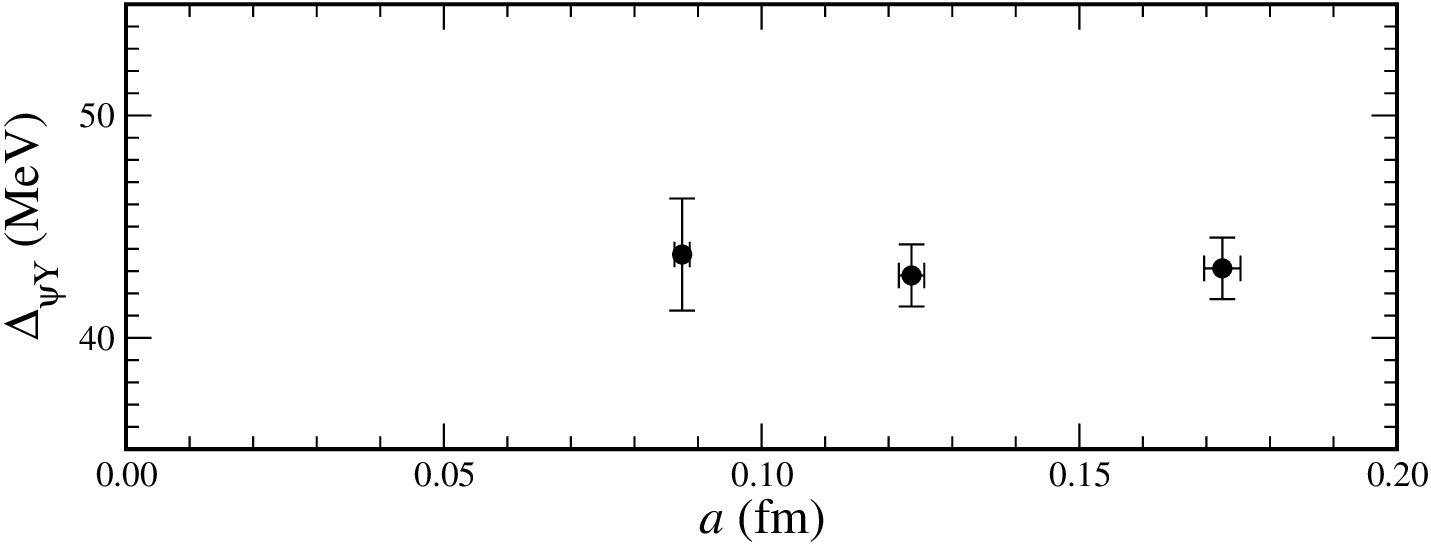}
	\caption{Lattice-spacing dependence of $\Delta_{\psi\Upsilon}$.}
	\label{fig:lattice}
	\vskip -1.0em
\end{figure}

Figure~\ref{fig:lattice} shows how $\Delta_{\psi\Upsilon}$ depends on
lattice spacing~$a$.
The change is insignificant.
Lattice spacing dependence stems from all parts of the lattice QCD
Lagrangian.
In our case, the heavy-quark discretization effects, especially for the 
$c$ quark, are expected to dominate.
Unfortunately, the dependence on $m_ca$ [of the coefficients in
Eq.~(\ref{eq:EFT})] does not provide a simple Ansatz for extrapolation.

We shall treat discretization errors with Eq.~(\ref{eq:EFT}), using
calculations of the short-distance mismatch and estimates of
the~$\mathcal{O}_n$.
This approach is itself uncertain, but it is preferable to ignoring the
issue.
The results of such an analysis are in given in
Table~\ref{tbl:discretization}, and the following paragraphs explain how
the entries are obtained.

As usual, we classify the operators~$\mathcal{O}_n$ in Eq.~(\ref{eq:EFT})
according to the power-counting scheme of NRQCD
(or, for $D_s$ and $B_s$ mesons, HQET).
Table~\ref{tbl:discretization} lists those of order $v^4$ in NRQCD;
\begin{table}[bp]
	\centering
	\caption{Estimated shifts (in $\text{MeV}$) of masses and 
	splittings $\Delta_{\psi\Upsilon}$ and $\Delta_{D_sB_s}$
	at $a=\frac{1}{8}~\text{fm}$.
	Entries show what should be added to the masses and
	splittings to compensate for discretization errors.
	Dots ($\cdots$) imply the entry is negligible.}
	\begin{tabular*}{\columnwidth}%
{c@{\extracolsep{\fill}}|*{7}{@{\extracolsep{\fill}}r}} \hline\hline
		~operator~ & $m_{B_c}$ & $\half\bar{m}_\psi$ & $\half m_\Upsilon$ &
		\multicolumn{1}{c}{$\Delta_{\psi\Upsilon}$\hspace*{-0.6em}} &
		\hspace*{1em}$\bar{m}_{D_s}$ & $\bar{m}_{B_s}$ &
		\multicolumn{1}{c}{$\Delta_{D_sB_s}$} \\
		\hline
		$\bm{\Sigma}\cdot\bm{B}$ & $-14$ & $0$ & $+3$ & $-17$ & 
			0 & 0 & $-14$~~~~ \\
		Darwin & $-3$ & $-3$ & $\mp1$ & $\pm1$ & 
			$-4$ & $\cdots$ & $+1$~~~~ \\
		$(\bm{D}^2)^2$ & $+34$ & $+10$ & $\pm3$ & $+24$ & 
			$\cdots$ & $\cdots$ & $+34$~~~~ \\
		$D_i^4$ & $+16$ & $+5$ &  $\pm2$ & $+11$ & 
			$\cdots$ & $\cdots$ & $+16$~~~~ \\
		Total & & & & $+18$ & & & $+37$~~~~ \\
		\hline\hline
	\end{tabular*}
	\label{tbl:discretization}
\end{table}
in HQET they are of order $1/m_Q^n$, $n=1,2,3,3$.
The spin-orbit interaction
$\bar{h}^\pm i\bm{\Sigma}\cdot(\bm{D}\times\bm{E})h^\pm$ is
omitted, because its matrix elements vanish in the S-wave states
considered here.

The contribution of the hyperfine interaction
$\bar{h}^{\pm}i\bm{\Sigma}\cdot\bm{B}h^{\pm}$
cancels for spin-averaged masses~$\bar{m}$, by
construction, but we must still estimate its effect on $m_\Upsilon$
and~$m_{B_c}$.
In the heavy-quark Lagrangians we are using, the hyperfine coupling is
correctly adjusted only at the tree level.
Indeed we find discrepancies in the hyperfine splittings
$m_{D_s^*}-m_{D_s}$ and $m_{J/\psi}-m_{\eta_c}$ for the $c$ quark and
$m_{B_s^*}-m_{B_s}$ for the $b$ quark.
The size of the discrepancy agrees with the expectation from the
one-loop mismatch in the coefficient.
The hyperfine entries for $m_\Upsilon$ and $m_{B_c}$ are obtained
by combining the coefficient mismatch with the computed hyperfine
splittings.

For $m_{B_c}$, $\half\bar{m}_\psi$ and $\half m_\Upsilon$,
the matrix elements of the Darwin term
$\bar{h}^\pm\bm{D}\cdot\bm{E}h^\pm$ and the relativistic corrections
$\bar{h}^\pm\left(\bm{D}^2\right)^2h^\pm$ and $\sum_{i=1}^3\bar{h}^\pm
D_i^4h^\pm$ are obtained from potential models.
For $\bar{m}_{D_s}$ and $\bar{m}_{B_s}$ we use HQET dimensional
analysis:
$\langle\bm{D}\cdot\bm{E}\rangle\sim\bar{\Lambda}^3$,
$\langle D^4\rangle\sim\bar{\Lambda}^4$,
with $\bar{\Lambda}=700~\text{MeV}$.
Next we multiply the estimated matrix elements $\langle\mathcal{O}_n\rangle$
with the mismatch coefficients~$f_n(m_Qa)$.
We have explicit tree-level calculations of them for the Fermilab
Lagrangian used for the $c$ quark.
For the $b$ quark the mismatch starts at order~$\alpha_s$, so we take
$f_n$ to be of order $\alpha_s$ with unknown sign.
The resulting shifts from the $c$ quark are larger, but their sign is
definite.

The entries in Table~\ref{tbl:discretization} for $(\bm{D}^2)^2$ and
$D_i^4$ are uncertain.
The cancellations across each row are reliable, but the overall
magnitude could be larger.
The same potential model suggests a shift in our $m_{h_c}-\bar{m}_\psi$
of about $-10~\text{MeV}$, consistent with the computed
discrepancy~\cite{Davies:2003ik,diPierro:2003bu}.
Thus, the charmonium spectrum suggests that the entries are reasonable.

Table~\ref{tbl:discretization} suggests that our results for $m_{B_c}$
will be too low, and that $m_{B_c}$ will be lower with the heavy-light
baseline than with the quarkonium baseline.
We could apply the shifts in Table~\ref{tbl:discretization} to our
lattice~QCD results.
Our aim, however, is to test lattice QCD.
Therefore, we treat these shifts not as corrections but as uncertainties.
Since we claim to know the sign in the important cases, the associated
error bars are asymmetric.
Repeating this analysis at other lattice spacings yields consistent 
error estimates.

After extrapolating the light quark mass and accumulating the other
systematic uncertainties we find (at $a=\frac{1}{8}~\text{fm}$)
\begin{eqnarray}
	\Delta_{\psi\Upsilon} & = &   39.8 \pm  3.8 \pm
		11.2^{+18}_{-\;\,0}~\text{MeV}, \\
	 \Delta_{D_sB_s}      & = & -\left[1238   \pm 30   \pm
		11^{+\;\,0}_{-37}\right]~\text{MeV},
\end{eqnarray}
where the uncertainties are, respectively, from
statistics (after extrapolating in $m_l/m_s$),
tuning of the heavy-quark masses, and 
heavy-quark discretization effects.
The results for $\Delta_{\psi\Upsilon}$ at $a=\frac{1}{11}$,
$\frac{2}{11}~\text{fm}$ are completely consistent.
For the $B_c$ mass we find
\begin{eqnarray}
	m_{B_c} & = & 6304 \pm\;\,4 \pm 11^{+18}_{-\;\,0}~\text{MeV},
		\label{eq:QQfinal} \\
	m_{B_c} & = & 6243 \pm   30 \pm 11^{+37}_{-\;\,0}~\text{MeV},
		\label{eq:sQfinal}
\end{eqnarray}
restoring, respectively, the quarkonium and heavy-quark baselines.
We have carried out more checks on the quark\-onium baseline,
so we take Eq.~(\ref{eq:QQfinal}) as our main result.
Given the rough nature of the last error bar, we consider
the agreement of the two results to be~reasonable.
Further work with more highly improved Lagrangians and at finer lattice
spacing should reduce this error.

\begin{figure}
	\centering
	\includegraphics[width=0.47\textwidth]{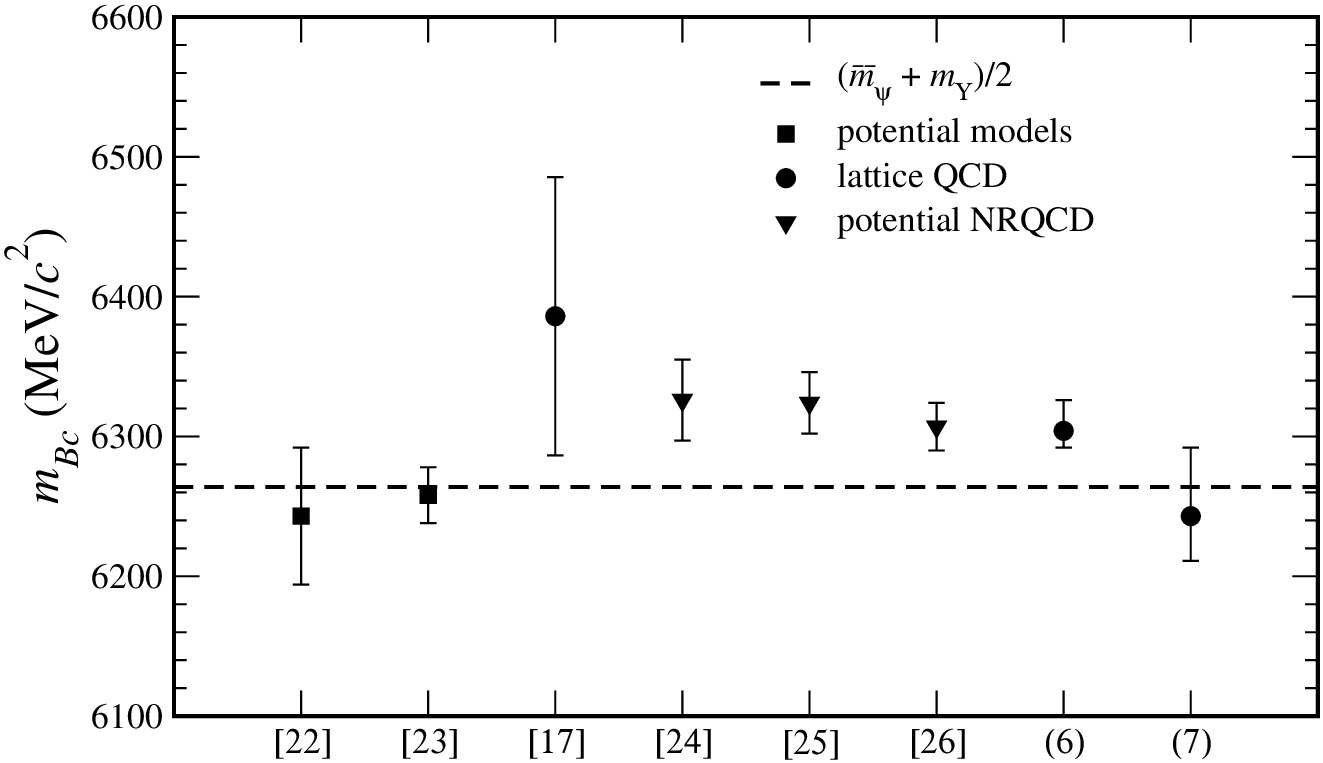}
	\caption{Comparison of theoretical work, with references in brackets
	and our equation numbers in parentheses.}
	\label{fig:compare}
\end{figure}

Our results are compared to other theoretical predictions in
Fig.~\ref{fig:compare}, including
potential models~\cite{Kwong:1990am,Eichten:1994gt},
quenched lattice QCD~\cite{Shanahan:1999mv}, and potential
NRQCD~\cite{Brambilla:2000db,Brambilla:2001fw,Brambilla:2001qk}.
The quarkonium baseline is shown for reference.
%
%
Our result is so much more accurate than the previous lattice QCD
result~\cite{Shanahan:1999mv}, simply because we have eliminated the
quenched approximation.
If our prediction, Eqs.~(\ref{eq:QQfinal}) and~(\ref{eq:sQfinal}), is
borne out by measurements, it lends confidence in lattice QCD,
not only in MILC's method for including sea quarks, but
also in the control of heavy-quark discretization effects using
effective field theory ideas.
Moreover, within this framework it is clear how to improve the lattice
QCD Lagrangian to reduce the remaining uncertainties.

We thank Junko Shigemitsu and Matthew Wingate for the $B_s$ masses used
here, and Estia Eichten and Chris Quigg for helpful discussions.
I.F.A. and C.T.H.D. are supported by the U.K. Particle Physics and
Astronomy Research Council.
A.G. is supported by the U.S.\ Department of Energy.
Fermilab is operated by Universities Research Association Inc.,
under contract with the U.S.\ Department of Energy.


\end{document}